\documentclass[11pt]{article}
\usepackage{comment}
\usepackage[margin=2.5cm]{geometry}
\usepackage{amsmath,amssymb,extarrows,mathtools,graphicx,subfigure,setspace,bbold}
\usepackage{cite}
\usepackage{slashed}
\usepackage[dvipsnames]{xcolor}
\usepackage{hyperref}
\hypersetup{colorlinks=true, linkcolor=blue, citecolor=magenta, linktoc=page}
\numberwithin{equation}{section}
\newcommand{\be}{\begin{equation}}
\newcommand{\bea}{\begin{eqnarray}}
\newcommand{\eea}{\end{eqnarray}}
\newcommand{\ba}{\begin{align}}
\newcommand{\ea}{\end{align}}
\newcommand{\ee}{\end{equation}}

\usepackage{chngcntr}
\counterwithout*{figure}{section}

\begin{document}

\begin{titlepage}
\vspace{10mm}


\vspace*{20mm}
\begin{center}

{\Large {\bf  Holographic Complexity of Subregions in the Hyperscaling Violating Theories }\\
}

\vspace*{15mm}
\vspace*{1mm}
{Z. Borvayeh${}^{a,\dagger}$,  M. Reza Tanhayi $^{b,c,\star}$ and 
S. Rafibakhsh${}^{ a}$}

 \vspace*{1cm}
{\it ${}^a$ Plasma Physics Research Center, Science and Research Branch, Islamic Azad University\\ Tehran 14665/678,  Iran\\
	${}^b$ Department of Physics, Faculty of Basic Science, Central Tehran Branch, Islamic Azad University,  Tehran, Iran  \\
\it ${}^c$ School of Physics,
Institute for Research in Fundamental Sciences (IPM)\\
P.O. Box 19395-5531, Tehran, Iran\\
}

 \vspace*{0.5cm}
{E-mails: {\tt $^\star$mtanhayi@ipm.ir, $^\dagger$zahra.borvayeh@srbiau.ac.ir}}%

\vspace*{1cm}
\end{center}

\begin{abstract}

In this paper, we use the complexity equals action proposal and investigate holographic complexity for hyperscaling violating theories on different subregions of space-time enclosed by the null boundaries. We are interested in computing the onshell action for certain subregions of the intersection between the Wheeler DeWitt  patch and the past, as well as, the future interior of a two-sided black brane. More precisely, we extend the results of Ref. \cite{Alishahiha:2018lfv} in parts, to hyperscaling violating geometries and to find the finite onshell action, we define the proper counter terms.  We show that in computing the rate of complexification the dynamical exponent plays a crucial rule, but, at the late time, rate of the complexity growth is independent of the hyperscaling parameters.

\end{abstract}

\end{titlepage}

\newpage
\tableofcontents
\noindent
\hrulefill
\onehalfspacing

\section{Introduction}

 $\,\,\,\,$In the context of black hole physics, recent progress indicates that there might be a connection between black hole physics and quantum information theory. Moreover, motivated by studying black holes, it is claimed that the compatibility of the laws of quantum mechanics and gravity demands that the universe might be considered as a hologram\cite{Moosa:2017yvt}. That means for a region of the universe all the information  should be encoded on a holographic screen that lives in one-lesser dimension. In fact, holography, as a powerful tool, helps us to study strongly interacting large $N$ quantum field theories   \cite{Brown:2015bva, Brown:2015lvg}. According to the holographic conjecture, a quantum field theory is mapped to the gravitational theory in one-higher dimensions, in a way that the field theory lives on the boundary of the space-time\cite{Ryu:2006ef}. This context is named as the holographic principle. 
In this way, Anti de Sitter / Conformal Field Theory (AdS/CFT) correspondence is an example where the holographic principle becomes manifest. This correspondence indicates that there is a deep relation between quantum information theory (e.g., entanglement and complexity) and  quantum gravity (e.g., area and volume). If this conjecture works, one may expect that, for instance, quantum information theory may play an important role in understanding the nature of space-time geometry \cite{Moosa:2017yvt,Tanhayi:2018gcj,Cottrell:2017ayj,Chapman:2018dem,Agon:2018zso}. 

On the other hand, besides the subjects say as horizon, black hole entropy and
information paradox, one of the key questions is understanding the physics behind the
horizon. The main motivation comes from this intuition that the quantum state of a black hole is somehow encoded in its interior geometry. It is argued that the entanglement may be the essence of space-time geometry and it can be a bridge  between quantum information theory and quantum gravity. So that, quantum information theory might be able to shed light on these subjects. Indeed, recent advances on black hole physics have opened up a possibility to connect quantum information theory and back hole physics \cite{Dong:2012se,Alishahiha:2012qu,Alishahiha:2018tep}. To understand this possible connection, the AdS/CFT correspondence has played rather an important role. In this context, one can say holographic entanglement entropy and computational complexity may be thought of as examples which could make this connection more concrete. In particular, holographic complexity, by definition, might be able to give some information about the physics behind the horizon\cite{Akhavan:2018wla,Alishahiha:2019cib}.  However, in the context of quantum field theory as well as information theory, the entanglement is by itself a difficult quantity to compute. But, the holographic techniques  make this possible connection to be
understandable\cite{Alishahiha:2018lfv,Agon:2018zso}.  It is expected that computational holographic complexity will be useful for understanding the physics of black holes, holographic property of gravity, and especially Hawking radiation \cite{Moosa:2017yvt,Akhavan:2018wla,Alishahiha:2019cib}.

 In this paper, we use the AdS/CFT correspondence as a tool to calculate the computational complexity of a two-sided black brane in hyperscaling violating backgrounds. On the other hand, studying the time evolution of the computational complexity in hyperscaling violating backgrounds allows us to check the validity of various recent conjectures involving black holes and complexity.

As a matter of fact, the holographic complexity might be considered as a measure of how difficult it is to implement some unitary operations during computation\cite{Cottrell:2017ayj}. In quantum circuits, complexity is defined as the minimal number of gates used for processing the unitary operation.
In the context of AdS/CFT correspondence, the growth of the interior of a two-sided black hole from the CFT is perspective described by the complexity. More precisely, it has been proposed that the growth of the interior of the black hole is dual to the growth of the complexity of the dual CFT state \cite{Lehner:2016vdi,Cai:2016xho}, where it was conjectured that the complexity of the boundary state at a given time $t$ is proportional to the value of the onshell gravitational action $\mathcal{A}(t)$ of a certain bulk region\cite{Hashimoto:2017fga,Jefferson:2017sdb,Chapman:2017rqy,Yang:2017nfn,Khan:2018rzm,Hackl:2018ptj,Alves:2018qfv,Bhattacharyya:2018wym,Guo:2018kzl,Bhattacharyya:2018bbv}\footnote{Actually, there is another holographic proposal for computing the complexity of the boundary states made by Susskind and collaborators \cite{Stanford:2014jda, Susskind:2014rva, Susskind:2014yaa}. The proposal states that the complexity is dual to the codimension-one volume of the maximal spacelike slice anchored at the two given
	boundary times. This is named as complexity=volume conjecture and motivated in several works (see for example Ref.s \cite{Alishahiha:2015rta,Momeni:2016qfv, Mazhari:2016yng}).}. This bulk region is the dependence domain of a Cauchy slice anchored on the boundary at time $t$. This conjecture is known as complexity equals action ($\mathcal{C}\mathcal{A}$) conjecture and the bulk region is called the Wheeler DeWitt (WDW) patch. The $\mathcal{C}\mathcal{A}$ conjecture is defined by \cite{Brown:2015bva,Brown:2015lvg,Cottrell:2017ayj}
\be
\mathcal{C}=\frac{\mathcal{A}_{\rm WDW}}{\pi}.\label{111}
\ee
On the other hand the rate of computation by the system is bounded by the energy of the system. This limitation is a universal bound, known as Lloyd's bound \cite{{Lloyd:2000}}, given by  
\be
\frac{d\mathcal{C}}{d\tau}\le\frac{2E}{\pi},
\ee
where $E$ is the average energy of the state at time $t$. One can use  $\mathcal{C}\mathcal{A}$ and also equation \eqref{111}, to find the rate of complexification at late times for isolated two-sided black holes.

On the other hand, in many condensed matter systems, at critical points the theory becomes conformally invariant. In these cases, when we rescale the spatial and temporal coordinates with a constant, the system stays invariant. However, there are systems that at their critical points do not scale as above. From holographic point of view such systems are dual to Lifshitz and hyperscaling violating geometries. 
The aim of this paper is to study complexity and subregion complexity in a wider family of states supporting
both anisotropic and also hyperscaling violating exponents\cite{Alishahiha:2012cm,Alishahiha:2019lng}. 

The layout of the paper is as follows. In the next section we briefly review the hyperscaling violating backgrounds and we also compute the onshell action on the WDW patch.  In
section 3, we use the onshell action to study the complexity in this background, actually in this section we almost review the computation of \cite{Alishahiha:2018tep} which we need them for the next section. Section 4, is devoted to the study of complexity of some certain subregions inside the black holes. This is done by computing the onshell action on the intersection of 
past and future interiors of the black brane with the WDW patch. In this section we also compute the
onshell action for past patch subregion in hyperscaling violating background. In fact, this special subregion is located between past singularity and continued past null boundaries. Finally, we present a discussion of our
results in section 5.

\section{Action on WDW Patch in Hyperscaling Violating Background}
Based on conjectured AdS/CFT proposal, AdS geometries are dual to the field theory with conformal symmetry. On the other hand, field theories which are scale-invariant but not conformal invariant are important, as long as, many physical systems in their critical points exhibit a rather different scaling in space and time and do not respect conformal invariance.  For example, in addressing the Landau-Fermi liquids, one needs Lifshitz type metrics in dual gravity theory  where the spatial and time coordinates of the field theory have been scaled differently. Therefore investigating the holographic dual models for such systems seems to be important. \\
In the theory with the Lifshitz fixed point, space and time scale differently as following\cite{Alishahiha:2018tep,Taylor:2015glc}
\be \label{13}
t\rightarrow\zeta^z t \;,\; x_i\rightarrow \zeta x_i\;,\; r\rightarrow \zeta r,
\ee
where $z$ is known as the dynamical critical exponent that equals to one in conformal field theories.
Simply, a Lifshitz invariant theory is spatially isotropic and homogeneous and admits the non-relativistic scaling symmetry \eqref{13}.
Moreover, a full class of scaling metrics can be obtained by considering both dilaton scalar field and an abelian gauge field and the resultant geometry named as the hyperscaling violating geometries. The corresponding action is given by\cite{Alishahiha:2012qu} 
\begin{equation}\label{2.1}
{\cal {A = }}  \frac{1}{{16\pi {{G}_N}}}\int {{{\rm{d}}^{{\rm{d}} + 2}}{\rm{x}}\sqrt { - g}} \left[ {R - \frac{1}{2}{{\left( {\partial \phi } \right)}^2} + V(\phi) - \frac{1}{4}{{e}^{\eta \phi }}}{(F_{\mu\nu})^2}  \right],
\end{equation}
 $G_N$ is the Newton constant, the potential of the scalar field and the vector field are given by
\begin{equation}\label{2.3}
V(\phi)={{\rm{V}}_0}{{\rm{e}}^{\xi \phi }},
 \;\;\;\;\;A^t=\frac{L}{r_f^{\frac{\theta}{d}}}\sqrt{\frac{2(z-1)}{d+z-\theta}}\frac{1}{r^{d+z-\theta}},\;\;\;\;e^{-\phi}=r^{\sqrt{2(d-\theta)(z-1-\theta)}},
\end{equation}
 where $L$ is the radius of the geometry and $r_f$ is a dynamical scale where the metric may not be a good description for UV complete theory above it\cite{Dong:2012se}.  In the above equations,  $\eta$, $\xi$ and $V_0$ are free parameters of the model defined by 
\bea\;\;\;
\;\;\;\;\;\;\;\;V_0&=&\frac{r_f^{\frac{2\theta}{d}}}{L^2}\Big(d+z-\theta-1\Big)\Big(d+z-\theta\Big),\,\,\,
 \xi=\frac{2\theta}{d\sqrt{2(d-\theta)(z-1-\theta)}},\nonumber\\ \eta&=&\frac{2\theta(d-1)-2d^2}{\sqrt{2(d-\theta)(z-1-\theta)}},
\eea
where the vector field produces
an anisotropy of the theory while non-trivial scalar potential leads to hyperscaling violating
factor. It is worth mentioning that in this theory, it is useful to define an effective dimension $d_e=d-\theta$, an effective hyperscaling violating exponent $\theta_e=\frac{\theta}{d}$ and also an effective scale $L_e=\frac{L}{r_f^{\theta_e}}$. Throughout this paper we set $L_e=1$. 
The solutions are given by  \cite{Taylor:2015glc, Alishahiha:2012qu} 
\be\label{metric}
ds^2=\frac{1}{r^{2(1-\theta_e)}}
\left(\frac{-f(r)}{r^{2(z-1)}}dt^2
+\frac{dr^2}{f(r)}+\sum_{i=1}^d d\vec{x}^2\right),
\ee
the blacking function $f(r)$ is also given by
\be
f(r)=1-\left(\frac{r}{r_h}\right)^{d+z-\theta},
\ee
where $r_h$ is radius of horizon. The Hawking temperature and the entropy are as follows 
\be
T=\frac{d+z-\theta}{4\pi r_h},\,\,\,\,\,\,\,\,\,S_{th}=\frac{V_d}{4G_Nr_h^{d-\theta}}\,.
\ee
Note that in the above equations $\theta$ is the hyperscaling violation exponent and $V_d$  stands for the volume corresponding to the space parametrized by the coordinates  
$x_i, \,i=1,\cdots d$.
This metric is not scale invariant, but under the scale transformation \eqref{13}, transforms as
\be
ds\rightarrow\zeta^{\theta_e} ds.
\ee
It should be mentioned that from null energy condition, one might write\cite{Alishahiha:2012qu,Dong:2012se}
\be
(d-\theta)\Big(d(z-1)-\theta\Big)\ge0,\;\;\;\;
(z-1)(d+z-\theta)\ge0.
\ee
In the following we suppose $d>\theta$ which results in $z\ge1$.\footnote{Note that there is another possibility of $\theta>d$ where this leads to the solution which is unstable\cite{Dong:2012se}.}
It is straightforward to see that the action density might be written as \cite{Lehner:2016vdi}
\be\label{bulk}
\sqrt{-g}\left(R-\frac{1}{2}(\partial\phi)^2+V_0 e^{\xi\phi}-\frac{1}{4} e^{\eta\phi}F^2\right)=-2(1-\theta_e)(d_e+z)\frac{1}{r^{d_e+z+1}}.
\ee
\begin{figure}
	\begin{center}
		\includegraphics[scale=0.85]{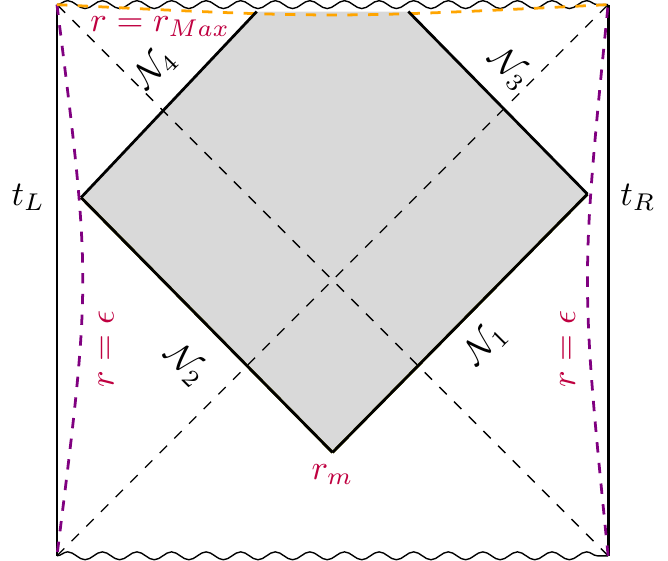}
		\end{center}
	\caption{Penrose diagram of the WDW patch of an eternal AdS black hole. ${\cal N}_i$ are null boundaries and it is supposed that $t_R=t_L$. 
		To find the complexity, the onshell action should be computed on this patch.}
	\label{WDWP1}
\end{figure}

According to `Complexity = Action' proposal, one needs to evaluate the onshell action inside the WDW patch as shown in Fig.\ref{WDWP1}. The WDW patch contains space-like and time-like boundaries, so that, it is known that the complete action should have certain Gibbons-Hawking terms defined at those boundaries. Moreover, the null boundaries as well as the joint points (points of intersection of these null boundaries with any other boundary) have their own stories and it is crucial to add the corresponding Gibbons-Hawking terms as well as certain joint actions. According to the well-defined variational principle, one can write the following action \cite{Alishahiha:2018tep}
\bea\label{ACT0}
\mathcal{A}^{(0)}&=&\frac{1}{16\pi G_N}\int d^{d+2}x\sqrt{-g}\Big(R-\frac{1}{2}(\partial\phi)^2+V_0 e^{\xi \phi}-\frac{1}{4} e^{\eta\phi}F^2\Big)+\frac{1}{8\pi G_N}
\int_{\Sigma^{d+1}_t} K_t\; d\Sigma_t\cr &&\cr
&& \pm\frac{1}{8\pi G_N} \int_{\Sigma^{d+1}_s} K_s\; d\Sigma_s\pm \frac{1}{8\pi G_N} 
\int_{\Sigma^{d+1}_n} K_n\; dS 
d\lambda \pm\frac{1}{8\pi G_N} \int_{J^d} a\; dS\,.
\eea
Here,  $\lambda$ is the null coordinate which is defined on 
the null segments; the time-like, space-like, and null boundaries and also joint points are denoted by $
\Sigma_t^{d+1}, \Sigma_s^{d+1}, \Sigma_n^{d+1}$ and $J^d$, respectively. The extrinsic 
curvatures of the corresponding boundaries are given by $K_t, K_s$ and $K_n$. On the other hand at the intersection of the boundaries, the function $a$ is defined by the logarithm of the inner product of the 
corresponding normal vectors. It is worth to mention that the relative position 
of the boundaries and the bulk region of interest identify the sign of different terms in the above action \cite{Lehner:2016vdi}. 

In this paper we are interested in computing the onshell action  in hyperscaling violating backgrounds for
the interior region of an eternal static neutral black brane in the generic dimensions which are dual
to a thermal state on the boundary.


\section{Complexity in Hyperscaling Violating Background }

 In this section, by making use of $\mathcal{C}\mathcal{A}$  proposal, we briefly review some recent results of the complexity for a two-sided black brane in hyperscaling violating backgrounds\footnote{More details can be found in Ref.\cite{Alishahiha:2018tep}, see also Ref.s\cite{Momeni:2017kbs, Momeni}. }. According to this proposal, the onshell action on WDW patch should be computed as shown in Fig.\ref{WDWP1}. However, the complete action is needed which contains the boundary terms. The crucial fact is that one needs some proper terms to add to the action of Eq. \eqref{ACT0} as it is not invariant under a reparametrization of the
null generators. Therefore,  in order to maintain the invariance under a reparametrization of the null
generators, an extra term is needed which should be added to the action. In Ref.\cite{Lehner:2016vdi}, it has been shown that such term might be given by
\be\label{AMB}
\mathcal{A}^{\rm amb}=\frac{1}{8\pi G_N}\int_{\Sigma_n^{d+1}} d^dxd\lambda\,\sqrt{\gamma}\,\Theta\,
\log\frac{|\Theta|}{d_e},
\ee
in which $\gamma$ is the determinant of the induced 
metric on the joint point, that by definition, two null segments intersect. $\Theta$ is defined by and $\Theta$ is defined by
\be
\Theta=\frac{1}{\sqrt{\gamma}}\frac{\partial\sqrt{\gamma}}{\partial\lambda}.
\ee 
Therefore, up to this level, the onshell action is  $\mathcal{A}=\mathcal{A}^{(0)}+\mathcal{A}^{\rm amb}$.
The symmetry of the  Penrose diagram in Fig.\ref{WDWP1}, demands that a symmetric 
configuration with times $t_R=t_L=\frac{\tau}{2}$ should be considered.
According to Fig.\ref{WDWP1}, it is obvious that there are four null boundaries of the corresponding WDW patch which are given by  
\bea
&&{\cal N}_1:\,\,t=t_R-r^*(\epsilon)+r^*(r),\;\;\;\;\;\;\;\;\;\;{\cal N}_2:\,\,t=-t_L+r^*(\epsilon)-r^*(r),\cr &&\cr
&&{\cal N}_3:\,\,t=t_R+r^*(\epsilon)-r^*(r),\;\;\;\;\;\;\;\;\;\;{\cal N}_4:\,\,t=-t_L-r^*(\epsilon)+r^*(r),
\eea
and also the location of the joint point $m$  is given by (note that in our notation we have used  $r^*(r)\leq 0$)
\be\label{rm}
\tau\equiv t_L+t_R=2(r^*(\epsilon)-r^*(r_m)).
\ee
Now, we want to compute the onshell action over the corresponding WDW patch. In fact the action has several parts including bulk, boundaries  and joint
terms. Making use of  Eq.\eqref{bulk}, the bulk action becomes
\bea
\mathcal{A}^{\rm bulk}_{\rm WDW}= -\frac{V_d}{4\pi G_N}(1-\theta_e)(d_e+z) \bigg(2\int_\epsilon^{r_{Max}} 
\frac{dr}{r^{d_e+z+1}}
(r^*(\epsilon)-r^*(r))\nonumber\\ +\int_{r_m}^{r_{Max}} \frac{dr}{r^{d_e+z+1}} 
(\frac{\tau}{2}-r^*(\epsilon)+r^*(r))\bigg)\,.
\eea
By integrating by parts and making use the following relation  $$r^*(r)=\int \frac{dr}{r^{1-z}f(r)},$$ the bulk action can be rewritten as follows 
\bea
 \mathcal{A}^{\rm bulk}_{\rm WDW}&=&-\frac{V_d}{4\pi G_N}(1-\theta_e)\left(\frac{(\tau+\tau_c)}{2 r_h^{d_e+z}}+\frac{2}{d_e\epsilon^{d_e}}-\frac{1}{d_e r_m^{d_e}}\right).
\eea
Note that the critical time is defined by $\tau_c=2(r^*(\epsilon)-r^*(r_{Max}))$ where the time derivative 
of complexity vanishes below $\tau_c$. \\
The next step is to compute the contribution of the  boundaries. However, for the null directions after making use the affine parametrization, one can show that the corresponding boundary terms vanish. In addition, it is crucial to mention that for the null directions after affine parametrization, the extrinsic curvature vanishes, so that, the null boundaries have no contribution. Therefore, one should only consider the space-like boundary at future singularity and the corresponding contribution is given by
\be\label{surf1}
\mathcal{A}^{\rm surf}_{\rm WDW}=-\frac{1}{8\pi G_N}\int d^dx\, \int_{-t_L-r^*(\epsilon)+r^*(r)}^{
t_R+r^*(\epsilon)-r^*(r)} dt\, \sqrt{h}K_s 
\Big|_{r=r_{\rm Max}},
\ee
where $h$ is the determinant of the induced metric and $K_s$ is the trace of extrinsic curvature of the boundary at $r=r_{\rm Max}$. In order to compute this term, we note that for a constant  surface $r$, after making use of metric \eqref{metric}, one might write
\be
\sqrt{h} K=-\sqrt{g^{rr}}\partial_r\sqrt{h}=-\frac{1}{2}\frac{1}{r^{d_e+z-1}}\left(\partial_r f(r)-\frac{2(d_e+z-\theta_e)}{r}f(r)\right).
\ee
Inserting  the above expression into Eq.\eqref{surf1} for $r=r_{\rm Max}$  one obtains   
\bea\label{SURF}
\mathcal{A}^{\rm surf}_{\rm WDW}=\frac{V_d}{8\pi G_N }(d_e+z-2\theta_e)\,\frac{\tau+\tau_c}{2r_h^{d_e+z}}\,.
\eea
There are also five joint points: two joint points at the future singularity and three at $r=\epsilon$ and $r=r_m$. The first two points have no contributions, on the other hand for the three remaining points, the contributions are evaluated as follows 
\bea
\mathcal{A}^{\rm joint}_{\rm WDW}= \frac{-1}{4\pi G_N}\int_{\epsilon} d^dx \sqrt{\gamma} \;
\log \frac{\left|{k_1\cdot k_2}\right|}{2}
+\frac{1}{8\pi G_N}\int_{r_m} d^dx \sqrt{\gamma} \;\log \frac{ \left|k_1\cdot k_2\right|}{2}\,.
\eea
Note that at $r=\epsilon$ there are two joint points at the  left and right boundaries. The null vectors are denoted by $k_1$ and $k_2$ and in the hyperscaling violating background, they are given by \cite{Alishahiha:2018tep}
\be
k_1^a={\alpha}\left(\frac{r^{2(z-\theta_e)}}{f(r)}(\partial t)^a+
r^{z-2\theta_e+1}(\partial r)^a\right),\;\;
\;\;\;\;k_2^a={\beta}\left(-\frac{r^{2(z-\theta_e)}}{f(r)}(\partial t)^a+
r^{z-2\theta_e+1}(\partial r)^a\right),
\ee
noting that one receives an ambiguity  in the normalization of normal vectors and two constants  $\alpha$ and $\beta$ are appearing due to this ambiguity. After doing some calculation, one obtains
\be
\mathcal{A}^{\rm joint}_{\rm WDW}=-\frac{V_d }{4 \pi G_N}\frac{\log(\alpha\beta  \epsilon^{2(z-\theta_e)})}{\epsilon^{d_e}}+\frac{V_d}{8 \pi G_N}
\left [\frac{\log(\alpha\beta r_m^{2(z-\theta_e)})}{r_m^{d_e}}-\frac{\log f(r_m)}{r_m^{d_e}}\right]\,.
\ee
The above expression has an ambiguity and should be fixed. This is done by adding another term given by Eq.\eqref{AMB}. In the present case for four null boundaries  in the hyperscaling violating background, we obtain 
\be
\mathcal{A}^{\rm amb}_{\rm WDW}=\frac{V_d }{4 \pi G_N}\left [\frac{\log(\alpha\beta\epsilon^{2(z-2\theta_e)})}{\epsilon^{d_e}}+\frac{2(z-2\theta_e)}{d_e\epsilon^{d_e}}\right]
-\frac{V_d}{8 \pi G_N} \left [\frac{\log(\alpha\beta r_m^{2(z-2\theta_e)})}{r_m^{d_e}}+\frac{2(z-2\theta_e)}{d_e r_m^{d_e}}\right].
\ee
On the other hand for the boundary corresponding to  $k_1$, one finds
	$$\frac{dr}{d\lambda}=\alpha r^{z+1-2\theta_e}\,\,\,\,\,\mbox{and}\,\,\,\,\,\,\, \Theta=-\alpha {d_e} {r^{z-2\theta_e}}.$$
At this stage, one has all components of the action on WDW patch which are needed to compute the complexity. Gathering all of them, one obtains 
 \bea\label{com}
\mathcal{\tilde{A}}_{\rm WDW}&&=\mathcal{A}_{\rm WDW}^{\rm bulk}+\mathcal{A}_{\rm WDW}^{\rm surf}+\mathcal{A}_{\rm WDW}^{\rm joint}+\mathcal{A}_{\rm WDW}^{\rm amb}\cr &&\cr
&&=\frac{V_d }{8\pi G_N}\bigg[
(d_e+z-2)\frac{(\tau+\tau_c)}{2 r_h^{d_e+z}}-\frac{\log |f(r_m)|}{r_m^{d_e}}
+\frac{2(z-\theta_e -1)}{d_e}\left (\frac{2}{\epsilon^{d_e}}-\frac{1}{r_m^{d_e}}\right)\bigg].
\eea
The resultant onshell action is UV-divergent and here, we follow the Ref. \cite{ Alishahiha:2019lng} to introduce a proper counter term to the action which is given by 
	\bea\label{coun1}
	\mathcal{A}^{\rm ct}= \frac{1}{8\pi G_N}\int d\lambda d^dx\, \sqrt {\gamma}\, \Theta \left (\frac{1}{2}\xi \phi+\frac{z-1}{d_e}\right).
\eea
After doing some straightforward calculation, the counter term \eqref{coun1} leads to 
\bea
\mathcal{A}^{\rm ct}=-\frac{ V_d}{8\pi G_N}\left(\frac{2\log\epsilon^{-2\theta_e}}{\epsilon^{d_e}}-\frac{\log r_m^{-2\theta_e}}{r_m^{d_e}}+\frac{2(z-\theta_e-1)}{d_e}\left(\frac{2}{\epsilon^{d_e}}-\frac{1}{r_m^{d_e}}\right)\right).
\eea
Now putting all together one obtains
\bea
\mathcal{A}_{\rm WDW}&=&\mathcal{\tilde{A}}_{\rm WDW}+\mathcal{A}^{\rm ct}\nonumber\\
&=&\frac{V_d}{8\pi G_N}\Big(\frac{d_e+z-2}
{2r_h^{d_e+z}}(\tau+\tau_c)-
\frac{\log |f(r_m)|}{r_m^{d_e}}\Big)\label{ac}\,.
\eea

It is also
interesting to note that for $r_m\rightarrow r_{Max}$ where $\tau\rightarrow \tau_c$, one gets
\bea
\mathcal{A}_{\rm WDW}=\frac{V_d}{8\pi G_N}\frac{(d_e+z-2)}
{r_h^{d_e+z}}\tau_c \,.\label{wdw1}
\eea
On the other hand, the growth rate of the complexity is given by  
\bea\label{as}
\frac{d\mathcal{C}_{\rm WDW}}{d\tau}&=&\frac{1}{\pi}\frac{d\mathcal{A}_{\rm WDW}}{d\tau}\nonumber\\
&=&\frac{2M}{\pi}\left(\frac{d_e+z-1}{d_e}+\frac{1}{2}\tilde{f}(r_m)\log|f(r_m)|\right),
\eea
in which
\be
\tilde{f}(r)=(\frac{r_h}{r})^{d_e+z}-1,\;\;\;\;\;\;M=\frac{ V_d}{16\pi G_N}\frac{d_e}{r_h^{d_e+z}},
\ee
 $M$ stands for the mass which  is proportional to the energy of the black brane. It is indeed a parameter at which the complexity approaches to at late times, namely,
\be
M=\frac{d_e}{d_e+z-1}E.
\ee
Note that in obtaining Eq.\eqref{as}, we have used  $$\frac{dr^*(r_m)}{d\tau}=-\frac{1}{2}\,\,\,\,\,\mbox{and}\,\,\,\,\, \frac{dr_m}{d\tau}=\frac{1}{2}\frac{f(r_m)}{r_m^{z-1}}.$$  

In the next section, we use $\mathcal{C}\mathcal{A}$ proposal to evaluate the 
onshell action on subregions inside the black brane.  

\section{ Complexity of Subregions Inside the Black Hole }

\begin{figure}
\begin{center}
\includegraphics[scale=0.9]{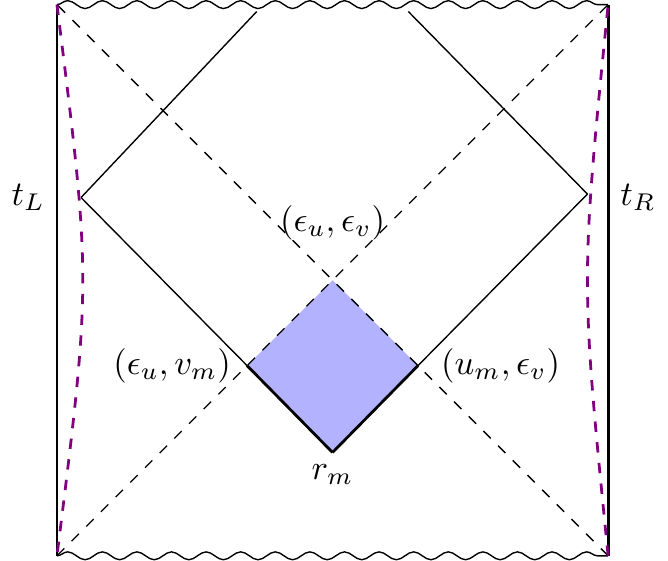}
\includegraphics[scale=0.9]{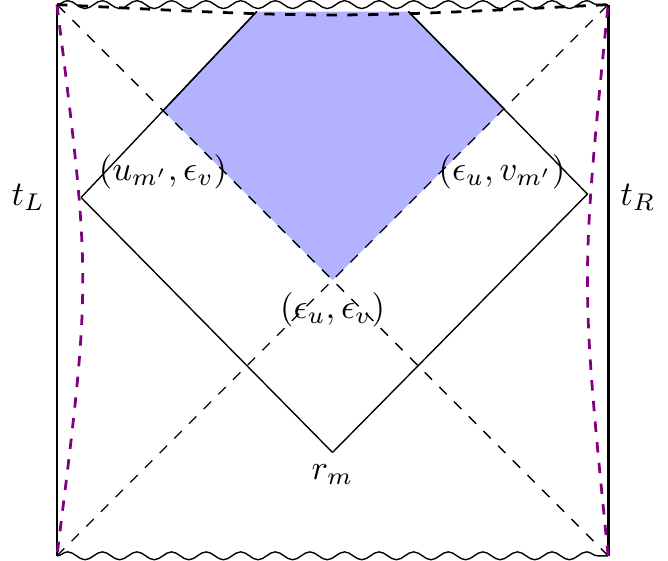}
\end{center}
\caption{ Intersection of WDW patch with the past ({\it left panel}) and future ({\it right panel}) interior of black brane. }
\label{fig:E}
\end{figure}

As already mentioned, the complexity is given by computing the onshell action 
over whole WDW patch, which was done in the pervious section. In order to get more insight of the interior of the black hole, we need to compute the onshell action in the intersection of WDW patch with the interior region of the black hole. As shown in figure \ref{fig:E}, the intersection consists of two main parts: past and future interiors. These parts are important, because, they play a crucial role in time dependency of the complexity of the dual state. In what follows, we consider the past and future interior of the black brane.


\subsection{Past interior of the black hole} 

Here, we are interested the onshell action in the intersection of the past interior of the black hole with the WdW patch as shown 
in the left panel of Fig.\ref{fig:E}. The results of the previous section can be used to compute different terms in writing the onshell action. First, let us consider the bulk term which reads as follows
 \bea
\mathcal{A}^{\rm bulk}_{\rm PI}&=&\frac{V_{d}}{4\pi G_N}(1-\theta_e)(d_e+z)\int_{r_h}^{r_m} \frac{dr}{r^{d_e+z+1}}
\int^{t_R-r^*(0)+r^*(r)}_{-t_L+r^*(0)-r^*(r)}dt\cr &&\cr
&=&\frac{V_{d}}{4\pi G_N}\frac{(1-\theta_e)}{d_e}\left(\frac{1}{\,r_m^{d_e}}-
\frac{1}{\, r_h^{d_e}}\right).
\eea
As it is clear there are four joint points, that one should consider their contribution in writing the onshell action; The first and the important one is located at $r=r_m$ and the remaining three points are located at $r=r_h$. To write the corresponding terms one might  define the following coordinate system  for the past interior \cite{Agon:2018zso}
\bea
u=-e^{-\frac{1}{2}f'(r_h)(r^*(r)-t)},\;\;\;\;\;\;\;\;\;\;\;\; v=-e^{-\frac{1}{2}f'(r_h)(r^*(r)+t)}\,.
\eea
Making use the above coordinate system, it is easy to check that the horizon is located 
at $uv=0$ meaning that $r^*(r_h)= -\infty$. The joint points which correspond to the solutions of this equation are, $u=0,v\neq 0$ and $u\neq 0,v=0$ and the third one is $u=0,v= 0$. It is important to know that at $r=r_h$  both $r^*(r)$  and 
$\log f(r)$ are singular. Therefore, the regularization is needed which may be done by setting the horizon at $u=\epsilon_u$ and $v=\epsilon_v$. As depicted in Fig.\ref{fig:E}, the three joint points are given by $(u_m,\epsilon_v)$ , $(\epsilon_u,v_m)$ and $(\epsilon_u,\epsilon_v)$, and the corresponding radial coordinates are denoted by $r_{v_m}, r_{u_m}$ and $r_\epsilon$, respectively. In this notation, one obtains the contribution of joint points as follows
\bea\label{joint-2} 
\mathcal{A}^{\rm joint}_{\rm PI}&=&\frac{V_d}{8\pi G_{N}}
\left(\frac{\log\frac{\alpha\beta  r_m^{2(z-\theta_e)}}{|f(r_m)|}}{r_m^{d_e}}
-\frac{\log\frac{\alpha\beta  r_{u_m}^{2(z-\theta_e)}}{|f(r_{u_m})|}}{r_{u_m}^{d_e}}
-\frac{\log\frac{\alpha\beta  r_{v_m}^{2(z-\theta_e)}}{|f(r_{v_m})|}}{r_{v_m}^{d_e}}
+\frac{\log\frac{\alpha\beta  r_\epsilon^{2(z-\theta_e)}}{|f(r_\epsilon)|}}{r_\epsilon^{d_e}}\right)\cr &&\cr
&=&-\frac{V_d}{8\pi G_{N}}
\bigg(\frac{\log{|f(r_m)|}}{r_m^{d_e}}
+\frac{\log{|f(r_{\epsilon})|}-\log{|f(r_{u_m})|}-\log{|f(r_{v_m})|}}{r_{h}^{d_e}}\cr &&\cr
&+&\frac{\log{\alpha\beta r_{h}^{2(z-\theta_e)}}}{r_{h}^{d_e}}
-\frac{\log{\alpha\beta  r_m^{2(z-\theta_e)}}}{r_m^{d_e}}\bigg),
\eea
where, we have used the limit $\{r_{u_m},r_{v_m},r_{\epsilon}\}\approx r_h$. Also one can use the following equation 
\be\label{Hlim}
\log |f(r_{u,v})|=\log|uv|+ c_0+{\cal O}(uv)\;\;\;\;\;\;\;\;\;\;\;{\rm for}\;\;uv\rightarrow 0,
\ee
to write the following relations \cite{Agon:2018zso} 
\bea
\log |f(r_{u_m})|&=&\log|u_m\epsilon_v|+ c_0+{\cal O}(\epsilon_v),\cr &&\cr
\log |f(r_{v_m})|&=&\log|\epsilon_u v_m|+ c_0+{\cal O}(\epsilon_u),\cr &&\cr
\log |f(r_{\epsilon})|&=&\log|\epsilon_u\epsilon_v|+ c_0+{\cal O}(\epsilon_u\epsilon_v).
\eea
Therefore, using the above relations, one can show that Eq.\eqref{joint-2}  simplifies as  follows
\be
\mathcal{A}^{\rm joint}_{\rm PI}=
-\frac{V_d }{8\pi G_{N}}
\left(\frac{\log{|f(r_m)|}}{r_m^{d_e}}
-\frac{\log (u_m v_m)+c_0}{r_{h}^{d_e}}
+\frac{\log{\alpha\beta r_{h}^{2(z-\theta_e)}}}{r_{h}^{d_e}}
-\frac{\log{\alpha\beta r_m^{2(z-\theta_e)}}}{r_m^{d_e}}\right),\label{amb}
\ee
where $c_0$ is a positive number. Again due to the normalization of the null vectors, there is an ambiguity in Eq.\eqref{amb} which can be removed 
by adding the extra term of Eq.\eqref{AMB}
to the action. Therefore, the following expression is obtained
\be
\mathcal{A}^{\rm amb}_{\rm PI}=
-\frac{V_d}{8\pi G_N}\left(\frac{\log\alpha\beta{ r_m^{2(z-2\theta_e)}}}
{r_m^{d_e}}+
\frac{2(z-2\theta_e)}
{{d_e}\,r_m^{d_e}}\right)+\frac{V_d}{8\pi G_N}\left(\frac{\log\alpha\beta { r_h^{2(z-2\theta_e)}}}
{r_h^{d_e}}+
\frac{2(z-2\theta_e)}
{{d_e}\,r_h^{d_e}}\right).
\ee
Now, gathering all the results, the onshell action in the subregion of past interior can be written as follows 
\bea
\mathcal{\tilde{A}}_{\rm PI}=\mathcal{A}_{\rm PI}^{\rm bulk}+\mathcal{A}_{\rm PI}^{\rm amb}+\mathcal{A}_{\rm PI}^{\rm joint}\hspace*{9cm}\cr &&\cr 
=\frac{V_d}{8\pi G_{N}}\left(\frac{\log|u_mv_m|+c_0}{r_h^{d_e}}+\frac{\log r_h^{-2\theta_e}}{r_h^{d_e}}-\frac{\log r_m^{-2\theta_e}}{r_m^{d_e}}+\frac{2(z-\theta_e-1)}{d_e}\left(\frac{1}{r_h^{d_e}}-\frac{1}{r_m^{d_e}}\right)\right),
\eea
in which, we have used 
the fact  that $$\log (u_m v_m)=-f'(r_h)r^*(r_m)=-\frac{(d_e+z)\tau}{2r_h^z}.$$ 
The corresponding counter term is given by 
\bea
\mathcal{A}_{PI}^{\rm ct}=\frac{ V_d}{8\pi G_N}\left[\frac{\log r_m^{-2\theta_e}}{r_m^{d_e}}-\frac{\log r_h^{-2\theta_e}}{r_h^{d_e}}+\Big(\frac{2(z-\theta_e-1)}{d_e}\Big)\left(\frac{1}{r_m^{d_e}}-\frac{1}{r_h^{d_e}}\right)\right].
\eea
Finally putting all together, one gets
\bea\label{amb1}
\mathcal{A}_{\rm PI}=\mathcal{\tilde{A}}_{\rm PI}+\mathcal{A}_{PI}^{\rm ct}
=\frac{ V_d}{8\pi G_N}\left(\frac{\log|u_m v_m|+c_0}{r_h^{d_e}}-\frac{\log|f(r_m)|}{r_m^{d_e}}\right)\nonumber\\
=\frac{ V_d}{8\pi G_N}\left(-\frac{(d_e+z)}{2r_h^{d_e+z}}\tau+\frac{c_0}{r_h^{d_e}}-\frac{\log|f(r_m)|}{r_m^{d_e}}\right).
\eea
We should mention that Eq.\eqref{amb1} depends on time trough its $r_m$ dependence. Therefore, for the past interior subregion the time derivative of the onshell action is given by
\bea
\frac{d\mathcal{C}_{\rm PI}}{d\tau}=\frac{1}{\pi}\frac{d\mathcal{A}_{\rm PI}}{d\tau}=\frac{M}{\pi}\tilde{f}(r_m)\log |f(r_m)|.
\eea

As a final remark note that for $r_m\rightarrow r_{\rm Max}$, the onshell action becomes as follows
\be
 \mathcal{A}_{\rm PI}=
\left(c_0-\frac{(d_e+z)\tau_c}{2r_h^z}\right)\frac{S_{th}}{2\pi}.
\ee

\subsection{Future interior of the black hole} 
In this subsection, we compute the onshell action for the future interior of the black hole. First, suppose we deal with the intersection of the WDW patch with the  
future interior of the black brane, as it is depicted in the right side of Fig.\ref{fig:E}. Making use the results of the previous subsection, the bulk term of the action becomes
\bea
\mathcal{A}^{\rm bulk}_{\rm FI}&=&-\frac{V_{d}}{4\pi G_N}(1-\theta_e)(d_e+z)\int_{r_h}^{r_{Max}} \frac{dr}{r^{d_e+z}}
\left(\frac{\tau}{2}+r^*(\epsilon)-r^*(r)\right)\cr &&\cr
&=&-\frac{V_{d}}{4\pi G_N}(1-\theta_e)\left(\frac{\tau+\tau_c}{2r_h^{d_e+z}}+\frac{1}{d_e\,r_h^{d_e}}\right).
\eea

It is clear from Fig.\ref{fig:E} that there are five joint points and should be considered. It can be shown that the contribution of two joint points at large $r_{Max}$ becomes zero, however, the other three joint points can be calculated as follows
\bea\label{joint-3} 
\mathcal{A}^{\rm joint}_{\rm FI}&=&\frac{V_d}{8\pi G_{N}}
\left(\frac{\log\frac{\alpha\beta  r_\epsilon^{2(z-\theta_e)}}{|f(r_\epsilon)|}}{r_\epsilon^{d_e}}
-\frac{\log\frac{\alpha\beta  r_{u_{m'}}^{2(z-\theta_e)}}{|f(r_{u_{m'}})|}}{r_{u_{m'}}^{d_e}}
-\frac{\log\frac{\alpha\beta  r_{v_{m'}}^{2(z-\theta_e)}}{|f(r_{v_{m'}})|}}{r_{v_{m'}}^{d_e}}
\right)\\&&\cr
&=&-\frac{V_d }{8\pi G_{N}}
\left(\frac{\log{|f(r_{\epsilon})|}-\log{|f(r_{u_{m'}})|}-\log{|f(r_{v_{m'}})|}}{r_{h}^{d_e}}
+\frac{\log{\alpha\beta  r_{h}^{2(z-\theta_e)}}}{r_{h}^{d_e}}
\right)\cr &&
\cr
&=&\frac{V_d }{8\pi G_{N}}
\left(\frac{\log |u_{m'} v_{m'}|+c_0}{r_{h}^{d_e}}
-\frac{\log{\alpha\beta  r_{h}^{2(z-\theta_e)}}}{r_{h}^{d_e}}
\right).\nonumber
\eea
Similar to the previous section for the null boundaries, making use of the affine parametrization one finds that the corresponding boundary terms have no contribution. Therefore, the surface term due to the future singularity is indeed the only term that one needs to compute, but this has been already computed in Eq.\eqref{SURF} and given by
 \be
\mathcal{A}^{\rm surf}_{\rm FI}=\frac{V_d}{8\pi G_N }(d_e+z-2\theta_e)\,\frac{\tau+\tau_c}{2r_h^{d_e+z}}\,.
\ee
Similar to the Eq.\eqref{AMB} one needs the following term in order to remove the ambiguity 
\be
\mathcal{A}^{\rm amb}_{\rm FI}=\frac{V_d}{8\pi G_N}\left(\frac{\log\alpha\beta r_h^{2(z-2\theta_e)}}
{r_h^{d_e}}+\frac{2(z-2\theta_e)}{d_e\,r_h^{d_e}}\right).
\ee
And the counter term is given by
\bea
\mathcal{A}_{FI}^{ct}=-\frac{V_e}{8\pi G_N}\left(\frac{\log r_h^{-2\theta_e}}{r_h^{d_e}}+\frac{2(z-\theta_e-1)}{d_e r_h^{d_e}}\right).
\eea
Finally, one obtains
\bea
\mathcal{A}_{FI}=\frac{V_d}{8\pi G_N}\left(\frac{d_e+z-2}{2r_h^{d_e+z}}(\tau+\tau_c)+\frac{\log|u_{m'}v_{m'}|+c_0}{r_h^{d_e}}\right)\nonumber\\
=\frac{V_d}{8\pi G_N}\left(\frac{(d_e+z-1)\tau}{r_h^{d_e+z}}+\frac{(d_e+z-2)\tau_c}{2r_h^{d_e+z}}+\frac{c_0}{r_h^{d_e}}\right)
\eea
The late time behavior for future interior part that is also given by\be
\frac{d\mathcal{C}_{\rm FI}}{d\tau}=\frac{1}{\pi}\frac{d\mathcal{A}_{\rm FI}}{d\tau}=\frac{2M}{\pi}\frac{d_e+z-1}{d_e}=\frac{2E}{\pi},
\ee
in which, we have used the following identity
$$\log |u_{m'}v_{m'}|=-{f'}(r_h)r^*(r_{m'})=\frac{d_e+z}{r_h}r^*(r_{m'})=\frac{(d_e+z)\tau}{2r_h^z}.$$

Before ending this section, let us mention a rather special subregion as shown by the colored
region in Fig.\ref{WDWP}. This subregion is important, because, it is indeed a part of space-time that causally connects to the operator behind the horizon localized at $r = r_m$. In other words, based on $\mathcal{CA}$ proposal, one can say that the onshell action in this subregion gives us the complexity associated with this operator. 

\subsection{Past patch}
In order to find the onshell action on this special region, it is noted that one may use  
the notation of the 
previous section and write the bulk term of the  onshell action as follows
\bea
\mathcal{A}^{\rm bulk}_{\rm past}&=&-\frac{V_d}{8 \pi G_N}(1-\theta_e)(d_e+z)\int^{r_{Max}}_{r_m} 
\frac{dr}{r^{d_e+z+1}}\int^{t_R-r^*(0)+r^*(r)}_{-t_L+r^*(0)-r^*(r)}dt\cr &&\cr
&=&-\frac{V_d}{4\pi
	G_N}(1-\theta_e)\left(\frac{\tau-\tau_c}{2r_h^{d_e+z}}-\frac{1}{d_e\,r_m^{d_e}}\right),
\eea
noting that in writing the second line, the integration by parts has been used. On the other hand, at the past singularity, the corresponding space-like  boundary term is given by 
\bea
\mathcal{A}^{\rm surf}_{\rm past}=
\frac{1}{8\pi G_N}\int d^dx\int^{t_R-r^*(0)+r^*(r)}_{-t_L+r^*(0)-r^*(r)}dt \sqrt{h}K_s\bigg|_{r=
	r_{\rm Max}}
=\frac{V_d}{8\pi G_N }
(d_e+z-2\theta_e)\frac{\tau-\tau_c}{2r_h^{d_e+z}}\,.
\eea
\begin{figure}
	\begin{center}
		\includegraphics[scale=0.85]{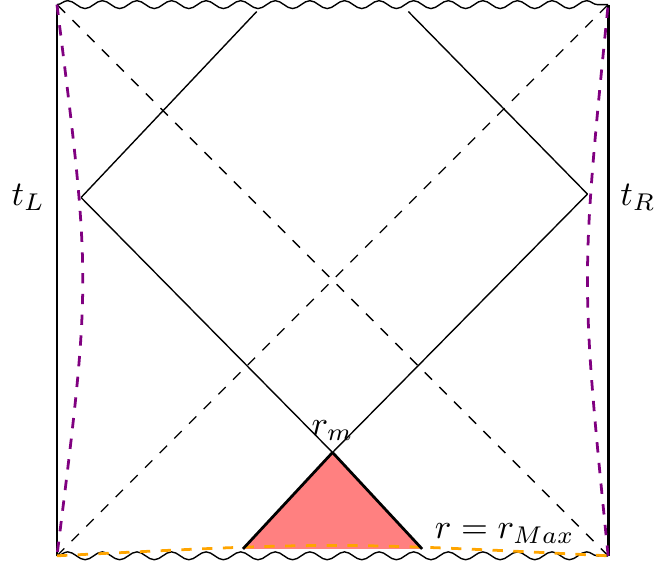}
	\end{center}
	\caption{Penrose diagram of the past patch.  
		}
	\label{WDWP}
\end{figure}
From Fig.\ref{WDWP}, it is clear that we have three  joint points; Two points at $r=r_{\rm Max}$ and one at $r=r_m$. It is worth noting that the contribution of joint points located at $r_{\rm Max}$ to onshell action would be vanished for large $r_{\rm Max}$, while the contribution of 
that at $r=r_m$ is given by  
\be 
\mathcal{A}^{\rm joint}_{\rm past}
=\frac{1}{8\pi G_N}\int d^{d-1}x \sqrt{\gamma} \;\log \frac{\left|{k_1\cdot k_2}\right|}{2}
=\frac{V_d}{8\pi G_N}\left(\frac{
	\log \alpha\beta  r_m^{2(z-\theta_e)}}{r_m^{d_e}}-\frac{\log |f(r_m)|}{r_m^{d_e}}\right).
\ee
Now, let us compute the last part namely in order to remove the ambiguity one should add a proper term to the onshell action.  Making use of the results of the previous section, one obtains  
\be
\mathcal{A}^{\rm amb}_{\rm past}
=-\frac{V_d}{8\pi G_N}\left(\frac{\log\alpha\beta r_m^{2(z-2\theta_e)}}{r_m^{d_e}}+
\frac{2(z-2\theta_e)}
{d_e\,r_m^{d_e}}\right),
\ee
also the counter term is
\bea
\mathcal{A}_{past}^{\rm ct}=\frac{V_d}{8\pi G_N}\left(\frac{\log r_m^{-2\theta_e}}{r_m^{d_e}}+\frac{2(z-\theta_e-1)}{d_e r_m^{d_e}}\right).
\eea
Gathering all the results, one arrives at
\bea
\mathcal{A}_{\rm past}=\frac{V_d}{8\pi G_N}\left(\frac{d_e+z-2}{2 r_h^{d_e+z}}\left(\tau-\tau_c \right)-\frac{\log|f(r_m)|}{r_m^{d_e}}\right).
\eea
We should mention that the onshell action for $r_m\rightarrow
r_{\rm Max}$ where $\tau\rightarrow \tau_c$, for past patch vanishes.

 
\section{Discussions}

In this paper, we used ``complexity equals action'' proposal, and studied the holographic complexity on the certain subregions enclosed by null boundaries including  the WDW patch for geometries with hyperscaling violating factor. We generalized the results of Ref.\cite{Alishahiha:2018lfv} for interior of the black hole, to hyperscaling violating geometries. It is worth mentioning that despite the fact that the solution and the rate of complexification depend on the dynamical exponent  $z$ and the hyperscaling violation exponent $\theta$, qualitatively, however, the rate of complexity growth behaves the same as that of Schwarzschild black brane. Moreover, we observed that at the late time, $\theta$ does not contribute. 

The growth rate of complexity suffers a logarithmic divergence goes as follows 
\bea
\frac{d\mathcal{C}_{\rm WDW}}{d\tau}
=\frac{2 E}{\pi}\left(1+\frac{d_e}{2(d_e+z-1)}\tilde{ f}(r_m) \log|f(r_m)| \right).\nonumber
\eea
Based on this result one can conclude that Lloyd's bound which is defined in terms of the mass of black brane is always violated for non trivial anisotropic and hyperscaling violating theories. This can be explained as follows: the value of $E$ which appears on the right hand side of Lloyd's inequality is always greater than (or equal to) the mass of the black brane. 

In principle, in computing the onshell action for a subregion one should take into account all terms in order to have a well-defined variational principle, moreover the crucial role of joint points and  time-like or space-like boundaries should be implemented. Particularly, we showed that the joint points play an important role in computing the rate of complexity. At late times, the past interior has no contribution to the rate of complexity where this rate approaches to $2M$ but violates the Lloyd's bound. This violation is due to the contribution of the joint point located at the past interior.

As a future work, it would be of
interest to investigate the effect of this kind of joint point in the late time behavior of the complexity for other geometries. Moreover, if the complexity played the role as an order parameter for phase transitions, it would be a relevant question that how does this parameter improve our knowledge about the role of hyperscaling violating parameters in the phase transitions?

\subsection*{Acknowledgments}
We would like to thank Mohsen Alishahiha for his very kind and generous support. We would like to acknowledge Amin Akhavan for his useful comments. We also thank, A. Naseh, F. Omidi for some related discussions. We also thank the referee for his/her comments.


\end{document}